\documentstyle[11pt,aaspp4]{article}

\def \ea{{\it et al.}}

\def \eg{{\it e.g.}}

\def\lsim{\mathrel{\rlap{\lower4pt\hbox{\hskip1pt$\sim$}}
    \raise1pt\hbox{$<$}}}
\def\gsim{\mathrel{\rlap{\lower4pt\hbox{\hskip1pt$\sim$}}
    \raise1pt\hbox{$>$}}}
\def \msol{\rm{M}$_\odot$}
\def \lsol{\rm{L}$_\odot$}
\def \mdot{\rm{M}$_\odot$~yr$^{-1}$}
\def \rsol{\rm{R}$_\odot$}
\def \kms{km~$\rm{s}^{-1}$}

\newcommand{\beq}{\begin{equation}}
\newcommand{\eeq}{\end{equation}}
\newcommand{\bdm}{\begin{displaymath}}
\newcommand{\edm}{\end{displaymath}}

\begin{document}

\title{Application of MHD Disk Wind Solutions to Planetary/Protoplanetary
Nebulae}

\author{A. Frank \altaffilmark{1} \&
E. G. Blackman\altaffilmark{1}}

\altaffiltext{1}{Dept. of Physics and Astronomy, University of
Rochester, Rochester, NY 14627-0171; afrank@pas.rochester.edu}

\begin{center}{\tt Submitted to Ap.J. \today}\end{center}

\begin{abstract}
Winds from accretion disks  have been proposed as the driving
source for precessing jets and extreme bipolar morphologies in
Planetary Nebulae (PNe) and proto-PNe (pPNe). Here we apply MHD
disk wind models to PNe and pPNe by estimating separately the
asymptotic MHD wind velocities and mass loss rates. We discuss
conditions which may occur in PNe and pPNe accretion disks that
form via binary interactions. We show that the resulting winds can
recover the observed momentum and energy input rates for PNe and
pPNe. High accretion rates ($M_a \approx 10^{-4}$ \mdot) may be
required in the latter case. We find that the observed {total}
energy and momentum in pPNe can be recovered with disk wind models
using existing disk formation scenarios.  When combined with
existing scenarios for accretion disk formation from disrupted
stellar companions, our models may provide an explanation for the
existence of high speed polar knots (FLIERS) observed in some PNe.
\end{abstract}


\section{INTRODUCTION}


\subsection{Introduction}

Planetary Nebulae (PNe) and proto-Planetary Nebulae (pPNe) are
believed to be the penultimate evolutionary stages of low and
intermediate mass stars ($M_* \le 8$ \msol). PNe and pPNe appear
on the sky as expanding plasma clouds surrounding a luminous
central star. As ground based telescopes increased their
resolution, elliptical and bipolar PNe were revealed (\cite{Balick87}).
The bipolar nebulae may be further
subdivided into ``butterfly," shapes in which the "waist" is
pinched at the central star, and ``bilobed'' PNe in which a pair
of larger outer lobes connects to a central and generally smaller
round or elliptical nebula (for a review see \cite{BalickFrank02}
and references therein). More recently, deeper and higher
resolution studies with the HST have shown evidence for narrow
collimated features that appear to be better described as {\it
jets} than {\it bipolar lobes}. The nature of these jets and other
highly collimated bipolar outflows in both PNe and pPNe remains a
subject of considerable debate (\cite{SahaiTrauger98}).

Considerable progress has been made in understanding the
hydrodynamic shaping of elliptical and bilobed PNe (for reviews
\cite{Frank99},\cite{BalickFrank02}), however the origin of
extreme butterfly nebulae as well as jets in PNe continues to pose
a number of problems for theorists.  The outstanding issues
include: (1) lack of a large scale torus to focus the outflows,
(2) the very large velocities in some jets, (3) point-symmetry of
outflows.

Beyond issues of morphology, there exists a formidable problem for
pPNe concerning the total momentum and energy in the outflows. A
number of observational studies have shown that radiatively
accelerated winds in pPNe cannot account for the high momentum and
energy implied by CO profiles.  This problem was identified first
by \cite{Knapp86}, and most recently and comprehensively
investigated by Bujarrabal, Alcolea and collaborators
(\cite{Bujarrabalea01}). The latter find that 21 of 23 CO-emitting
pPNe objects show outflows whose scalar momentum ($\Pi = M V$) are
more than $10^3$ times larger than that in the stellar radiation.
Thus $\Pi >> (L_*/c)\Delta t$, where here L$_*$ is the stellar
luminosity emitted during the pPNe outflow expansion lifetime
$\Delta t$. In light of these results both the {\it launching and
collimation} of winds in pPNe becomes problematic.

The dominant hydrodynamic theory for shaping PNe had been the
Generalized Interacting Stellar Winds (GISW) model
(\cite{Kwokea78}, \cite{Balick87}, \cite{Icke88}) in which a star
and its wind evolve from the AGB to a white dwarf. A slow, dense
(wind expelled during the AGB is followed by a fast, tenuous wind
driven off the contracting proto-white dwarf during the PNe phase.
Numerical models have shown this paradigm can embrace a wide
variety of nebular morphologies including {\em highly collimated
jets} (\cite{Ickeea92}, \cite{MellemaFrank97},
\cite{Borkowskiea97}) when the slow wind takes on aspherical
density distributions. While the GISW model {\it can} produce
narrow jets it usually requires a large-scale ``fat'' torus.  It
is difficult to imagine that a large-scale out-flowing gaseous
torus can provide a stiff precessing nozzle for production of
point symmetric flows (Recent results of Icke 2003 however
indicate that pure hydrodynamics may lead to point symmetric
shapes in some cases).

In addition, it is now recognized that fast ($\ge 100$ \kms)
bipolar outflows can occur in the pPNe or even the Post-AGB stage.
Objects like CRL 2688 and OH231.8+4.2 raise the question of how
high-velocity collimated flows occur when the star is still in a
cool giant or even supergiant stage (CRL 2688 has an F Supergiant
spectral type). Finally the GISW model assumes a radiation driven
wind.  As discussed above, the results of Bujarrabal \ea (2001)
make radiation driving untenable as a source for many pPNe flows.

Models invoking a toroidal magnetic field embedded in a normal
radiation driven stellar wind have shown considerable promise
(\cite{ChevalierLuo94}, \cite{RozyczkaFranco96},
\cite{GarciaSeguraea99}), \cite{GarciaSegura97}). Recent results,
however also imply that jets may form at smaller distances and
that such models may not begin with appropriate initial conditions
(\cite{GardinerFrank01}).  In addition, by their very nature, MWB
models can not account for the momentum excesses in pPNe since
they also require radiation driven winds. The fields are simply
too weak to power the observed outflows.

Thus there remains considerable uncertainty about the processes
which produce collimated jets in pPNe and PNe. Other systems which
produce jets such as YSOs, AGN and micro-quasars have been
modelled via a combination of magnetic and centrifugal forces from
accretion disks (\cite{BlanfordPayne82}, \cite{Shuea94},
\cite{PudritzKonigl00})

The success of these {\it Magneto-centrifugal Launching} (MCL)
models is such that it is worthwhile considering if such a
scenario can be applied to PNe and pPNe. Indeed, Morris (1987) and
Soker \& Livio (1994) mapped out scenarios in which accretion
disks form around binary PNe progenitors.  Each study equated the
existence of disks with the existence of jets. The details of the
jet launching and collimation mechanism where not, however,
specified. Recent work by Soker \& Rapport 2001, Soker 2001 and
\cite{SokerLivio01} have relied heavily on collimated winds from
disks but these works also do not specify how such winds are
launched or collimated. Thus application of MCL disk wind models
to PNe and pPNe would close an important gap in building a new
paradigm for these systems.  In particular MCL models may offer a
means of resolving issues associated with both PNe jet precession
(the underlying disk precesses via instabilities: Livio \& Pringle
(1996), \cite{Quillen01}) {\it and} those associated with pPNe
momentum excesses (no need for radiation pressure launching
winds).

Recent studies by Blackman, Welch \& Frank (2001) and Blackman \ea
~2001 have explored MCL paradigms for both the star and disk in
pPNe and PNe systems. They computed an upper limit on the magnetic
luminosity available to power an outflow, assuming that a dynamo
is the source of the large scale magnetic field. In this paper we
we provide further calculations along these lines deriving scaling
relations from the equations for MCL and separately estimate the
mass outflow rate and the asymptotic outflow velocity.  We then
compare to these results to PNe and recent observations of PNe.
When the dynamo is invoked to produce the field, the mechanical
wind luminosity and thus outflow rate are naturally linked to the
accretion rate.

In section 2 we derive the outflow speed and the mass loss rate by
combining results from magnetically driven wind theory and
dynamos. In section 3 we discuss models of pPNe and PNe disks. In
section 4 we apply the results of sections 2 and 3 to PNe and pPNe
and show that it is easy to solve the afore mentioned the pPNe
momentum and energy excess problems. In section 5 we conclude and
discuss open questions.

\section{Mass Outflow Rate and Asymptotic Wind Speed
from MCL Theory and Dynamo Theory}

\subsection{Magnetic Luminosity}
The basic physics of magneto-centrifugal launching of winds and
jets is well studied when a magnetic field distribution is imposed
on the disk (\cite{BlanfordPayne82}, \cite{Sakurai85}, \cite{PP92}
(hereafter PP92) \cite{Shuea94} \cite{Ostriker97},
\cite{PudritzKonigl00} (hereafter KP00)). These models {\it
presume} an initial field of a given strength and geometry, but
the tendency for large scale fields to diffuse (Lubow \& Pringle
1994; Blackman 2003) suggest that the field must be generated in
situ by a dynamo (e.g. Blackman et al 1999).  In this section we
briefly summarize how to combine the basics of Poynting flux
driven outflows with asymptotic wind solutions and mean field
dynamo theory to estimate the asymptotic wind speed and the
outflow accretion rate.

Magneto-centrifugal launching is a means of converting
gravitational binding energy in an accreting source into kinetic
energy of an outflowing wind.  The magnetic fields act as a drive
belt to extract angular momentum from the anchoring rotator and
launch the wind.  The magnetic luminosity, or equivalently, the
maximum magnetic power available for a wind can be obtained from
the integrated  Poynting flux (\cite{BFW01}). The field lines
rotate nearly rigidly at angular speeds associated with the
anchoring foot point $\Omega_0(r)$ up to the Alfv\'en radius
$r_A(r)$.  After this point the angular speed falls off with
$1/r^2$ (conserving specific angular momentum) and the field falls
off as $1/r$. The field is primarily poloidal out to the Alfv\'en
radius associated with each field line where the poloidal and
toroidal components are comparable. If the poloidal field falls
off as $1/r^2$ out to the Alfv\'en radius and is not too far from
the disk surface, the Poynting flux can be approximated by the
contribution from the Alfv\'en radius associated with field lines
anchored at the inner radius of the disk. That is, we have
\begin{eqnarray}
L_w & = & \frac{1}{2} {\dot M}_w u_{\infty}^2 \sim L_{mag} \equiv
\int ({\bf E} \times {\bf B}) \cdot d{\bf S}_A \sim
\int_{r_{i}}^{r_{A}(r_i)} \Omega (r)r B_pB_\phi rdr \sim  B_A^2
\Omega_0 r_A^3, \label{magicF}
\end{eqnarray}
where $r_0$ is the disk inner radius and where $B_A= B_\phi \sim
B_p$ at the Alfv\'en surface
 (i.e. the toroidal and poloidal field components are nearly
equal). Since the dominant contribution to the magnetic luminosity
comes from the field lines anchored at the innermost radius $r_i$,
in what follows {\it all quantities labeled with subscript $0$
refer to those values evaluated at $r=r_i$, the inner most disk
radius.} This also implies that $r_0=r_i$.

\noindent

\subsection{The Bernouilli Constant}
To estimate the mass outflow rate and outflow speed separately
a bit more work is required.
The MCL problem requires the construction of solutions for a
steady, ideal, isothermal magnetohydrodynamic flow. The isothermal
assumption eliminates the need for solving the energy equation,
but more complex assumptions can be used, \eg ~a polytropic law.
The system of equations to solve become mass conservation,
momentum conservation, and the steady state induction equation
(e.g. \cite{PP92}). Respectively,  these are
\begin{equation}
\nabla \cdot  (\rho {\bf u})  = 0
\label{2}
\end{equation}
\begin{equation}
\rho{\bf u} \cdot \nabla {\bf u}  =  \frac{1}{4\pi}(\nabla {\bf
\times} {\bf B}) {\bf \times} {\bf B} - \nabla P - \rho \nabla \Phi
\label{3}
\end{equation}
\begin{equation}
\nabla {\bf \times} ({\bf u} {\bf \times} {\bf B})    =  0
\label{primeq}
\end{equation}
where $\Phi$ is the gravitational potential due to the central
source. In cylindrically symmetric coordinates the physics of MCL
disk winds can be cast in axisymmetric form where the velocity and
magnetic fields are decomposed into toroidal ($\phi$) and poloidal
($r,z$) components,
\begin{eqnarray}
{\bf B} & = & {\bf B}_p + B_\phi {\hat \phi}\\
{\bf u} & = & {\bf u}_p + {\bf u}_\phi = {\bf u}_p + \Omega r~
{\hat \phi}
\label{vbrel}
\end{eqnarray}
where $\Omega$ is the rotational frequency of the plasma at a
point $(r,z)$.
Axisymmetry and $\nabla\cdot {\bf B}=0$ allow
${\bf B}_p$ to be expressed in terms of
a magnetic flux function, $a(r,z)$, such that
\begin{equation}
{\bf B}_p = \frac{1}{r} ({\bf \nabla} a ~ \times  \hat{\phi}).
\label{magflux}
\end{equation}
The magnetic surfaces, generated by rotation of a poloidal field
line about the axis, are surfaces of constant $a(r,z)$.

For axially symmetric configurations, (\ref{2}) and (\ref{primeq}),
imply that the poloidal velocity is always
parallel to magnetic surfaces, that is
\begin{equation}
\rho {\bf u}_p = k(a) {\bf B}_p, \label{parub}
\end{equation}
where the function $k$ is constant on a flux surface (KP00
equation 7). The induction equation and $\nabla \cdot {\bf B}=0$ also
imply
\begin{equation}
\rho u_{\phi} =  k(a) B_{\phi} + \rho r \Omega_o(a)   ,
\label{fer}
\end{equation}
where $\Omega_o(a)$ is a constant that is approximately equal to
the angular velocity in the disk where the magnetic surface is
tied (KP00 equation 9).

Using $\nabla\cdot{\bf B}=0$, Eqn. (\ref{2}), and
the azimuthal component of Eqn. (\ref{3}),
the conserved angular momentum per unit mass $l(a)$
becomes
\begin{equation}
l(a) = \Omega r^2 - \frac{r B_\phi}{4\pi k} = const(a)=\Omega_o r_A^2,
\label{amomc}
\end{equation}
where the last equality follows from using (\ref{fer})
and finding the value at the ``Alfv\'en radius''
$r_A(a)$  for a magnetic flux surface anchored to the disk at
radius  $r_o(a)$ (e.g. KP00 equation 11).
The Alfv\'en radius $r_A(a)$ defines the radial coordinate of the point
along a poloidal magnetic surface
when the  outflow speed on that surface equals the local poloidal Alfv\'en
velocity $u_A = B_p/\sqrt{4\pi \rho}$. The expression for
conservation of angular momentum (\ref{amomc})
thus relates contributions from the
angular momentum of matter and magnetic torques to a fiducial
value associated with the Alfv\'en point.

The poloidal component of Eqn. (\ref{3}) can be integrated
using $(\ref{fer})$
and (\ref{amomc}) to  give the total conserved
specific energy $U(a)$ carried by the wind
on magnetic surfaces in both kinetic energy and Poynting flux. By
defining $w$ as the enthalpy per unit mass,
the generalized Bernoulli integral then emerges (KP00 equation 14) as
\begin{equation}
\frac{1}{2}(u_p^2 + \Omega^2 r^2) + \Phi + w + \Omega_o(\Omega_o
r_A^2- \Omega r^2) = U(a) = const(a) \label{berneq}.
\end{equation}

\subsection{The Outflow Speed and Mass Outflow Rate}

We now assume a cold wind (such that $w$ can be ignored).
Because little acceleration occurs outside
the Alfv\'en surface we assume that
\beq
u_\infty= f u_A
\label{uinf}
\eeq
where $f\gsim 1$. We will solve for $u_A$ and constrain $f$ below.

To calculate $u_A$, we now solve the momentum equation for $r\le
r_A$  in the rotating frame. For a cold wind, the dominant force
components are the centrifugal force and the gravitational force.
For $r <r_A$, the magnetic field lines can be assumed to rotate
rigidly with the angular speed of their foot points at $r=r_0$. It
is straightforward  to see that the steady state radial momentum
equation can then be written (Blandford \& Payne (1982)) \beq {\bf
u}\cdot\nabla u_r = -\partial_r \Phi_{eff} = \left({GM_* \over
r_0}\right)\left({r \over r_0^2}-{rr_0\over (z^2 +
r^2)^{3/2}}\right). \label{bp} \eeq To keep the analysis as simple
as possible, consider the initial launch to be highly inclined,
such that $z << r$. Such an approximation is reasonable since
(\ref{bp}) implies that the field inclination from the disk must
make an angle $<\pi/4$ to the disk plane for launch. This result
for the field inclination follows from (\ref{bp}) by considering
$z << r$ and $(r-r_0) << r_0$ and carrying out an expansion of the
resulting force along the field line to second order in
$z^2/r_0^2$ and $(r-r_0)^2/r_0^2$. The reason for an outward force
is the gain in centrifugal force at $r>r_0$ from the fact that
``rigid'' field lines enforce co-rotation.  (Note in the following
sections we will use the fact that the dominant contribution to
the magnetic luminosity comes from the field lines anchored at the
innermost radius $r_i$ impling that $r_0=r_i$.)

Integrating (\ref{bp}) along the field line for the case $z<<r$
and taking the result at $r=r_A$ gives \beq u_A^2\simeq \Omega_0^2
r_A^2 \left(1+2{r_0^3\over r_A^3}\right) =q^2 \Omega_0^2 r_A^2,
\label{uaq} \eeq where we see that \beq 1< q^2=(1+2{r_0^3/
r_A^3})<3. \label{q} \eeq Now \beq u_A^2= B_A^2/4\pi \rho_A,
\label{alfven} \eeq where $\rho_A$ is the mass density at $r_A$,
and \beq \rho_A= {{\dot M}_{w,l}/\xi \over  r_A^2  u_A}= {{\dot
M}_w\over 4\pi  r_A^3 \Omega_0 q}, \label{rho} \eeq from  the mass
continuity equation, where $\xi$ is the solid angle $\le 4\pi$
corresponding to the launched outflow mass loss rate $M_{w,l}$
whereas $M_w=  4\pi M_{w,l}/\xi$ is the effective mass loss rate
were it quasi-spherical.

Using (\ref{uaq}), (\ref{alfven}) and (\ref{rho}), we then obtain

\beq {\dot M}_w={B_A^2 r_A\over q\Omega_0}. \label{mdot1} \eeq

But using (\ref{uinf}) and (\ref{magicF}) we get a separate
equation for ${\dot M}_w$, namely

\beq {\dot M}_w={2\over f^2 q^2}{B_A^2 r_A\over \Omega_0}.
\label{mdot2}
\eeq

By setting (\ref{mdot1}) equal to (\ref{mdot2}) we obtain $f^2 q=
2$. Thus from (\ref{q}) we must have

\beq
{2\over 3^{1/2}}<  f^2 <2.
\label{f}
\eeq

This is a narrow range of $f$ and for our crude order of magnitude
estimates we will take $f=1.2$, a value right in the middle of the
allowed range. For this choice of $f$, we obtain $q=1.4$.  For
simplicity, we will use these values in what follows. From
(\ref{q}) these values imply

\beq
r_A=1.27 r_0.
\label{ra2}
\eeq

Using these in (\ref{uinf}) and (\ref{mdot1})
then give
\beq
u_\infty \simeq  2.1\Omega_0 r_0.
\label{uinf2}
\eeq
and
\beq
{\dot M}_w\simeq 0.9 {B_A^2r_0\over \Omega}.
\label{mdot3}
\eeq
In the next subsection
we obtain an expression for $B_A$.

\subsection{Magnetic Field Strength}

Magnetic fields may form in these disks via dynamo processes
(\cite{RRS95}, BFW01). The topology of such a field, (the ratio of
poloidal $B_p$ and toroidal $B_\phi$ in the disk) and its
subsequent value in the coronae remains a subject of considerable
discourse (e.g. Blackman 2003). Here we assume that a field
produced by dynamos can drive a disk wind in the manner described
in the last section. This means that whatever combination of
toroidal and poloidal field is produced in the disk, we assume a
primarily polodial field in the corona where the wind launches.

Eqn. (\ref{primeq}) and $\nabla\cdot {\bf B}=0$ imply that the
radial magnetic field falls off with $r^2$ along the field line to
$r_A$. Thus \beq B_A^2 = B_0^2 (r_0/r_A)^4= 0.4 B_0^2, \label{ba}
\eeq where $B_0$ is the poloidal field at the disk surface and we
have used (\ref{ra2}). The square of the surface  poloidal field
$B_0^2$ can be estimated to be lower than the midplane poloidal
field squared $B_d^2$ by the density ratio to the $4/3$ power
(flux freezing). The density falls by a factor of $e^{-n}$, where
$n$ is the number of scale heights above the disk from where the
wind is launched. We then have \beq B_0^2 = e^{-4n/3} B_d^2 \sim
4\pi e^{-4n/3}\rho_d \alpha_{ss}^2c_s^2, \label{bd} \eeq where
$\alpha_{ss}$ is the disk viscosity parameter (Shakura \& Sunyaev
1973), $c_s$ is the sound speed, and $\rho_d$ is the disk density.
The latter similarity in (\ref{bd}) follows from estimating the
disk mean poloidal field from a helical mean field dynamo
(\cite{BFW01}, Blackman 2003). From mass conservation in the disk,
the  disk density satisfies \beq \rho_d = {{\dot M}_d \over 2 \pi
r_0 h_0 v_r }, \label{dd} \eeq where $h_0$ is the disk scale
height at $r=r_0$ and $v_r=\alpha_{ss} c_s h_0/r_0$ is the disk
radial accretion velocity, and ${\dot M}_d$ is the mass accretion
rate. Combining this with (\ref{ba}), (\ref{bd}) and (\ref{dd})
then implies \beq B_0^2 = {2 e^{-4n/3}\alpha_{ss} c_s {\dot
M}_d\over h_0^2} =2.5 B_A^2 . \label{bd2} \eeq


\subsection{Compiling the Formulae for Application to Observations}

Combining (\ref{bd2}) with (\ref{mdot3}) and using $c_s =\Omega
h_0$ for an accretion disk, we obtain \beq {\dot M}_w= 0.72
e^{-4n/3}\alpha_{ss}{r_0\over h_0}{\dot M}_d
\label{mdot4}
\eeq
for the mass loss rate. Combining this with (\ref{uinf2}) and (\ref{magicF})
gives
\beq
L_{mag}\simeq 1.6 e^{-4n/3}\alpha_{ss}{r_0\over h_0}{\dot M}_d \Omega_0^2 r_0^2.
\eeq
The
scaling for the momentum input $\dot{\Pi} = \dot{M}_w u_\infty$ of the
wind can then also be estimated via
\begin{equation}
\dot{\Pi} \sim L_{mag}/u_{\infty} \simeq 0.76
e^{-4n/3}\alpha_{ss}{r_0\over h_0}{\dot M}_d \Omega_0 r_0.
\end{equation}

In the above three relations, there is some freedom in choosing
$n$, the number of scale heights above the midplane from which the
wind launches (i.e. the location at which the plasma becomes
magnetically dominated).  One expects the corona to become
magnetically dominated after one or two scale heights thus $n=1.5$
is reasonable choice for the point at which we expect the ratio of
thermal to magnetic pressure ($\beta$) to make the transition
$\beta > 1$ to $\beta < 1$. Using this we obtain

\beq {\dot M}_w\sim 0.1 \alpha_{ss}{r_0\over h_0}{\dot M}_d,
\label{mdot5}
\eeq

\beq L_{mag}\simeq 0.22 \alpha_{ss}{r_0\over h_0}{\dot M}_d
\Omega_0^2 r_0^2, \label{maglum}
\eeq

and

\beq \dot{\Pi} \sim L_{mag}/u_{\infty} \simeq 0.14
\alpha_{ss}{r_0\over h_0}{\dot M}_d \Omega_0 r_0. \label{momentum}
\eeq

The last two expressions give estimates of the rate that energy
and momentum are input by MCL disk winds to the ambient medium,
and (\ref{uinf2}) gives the asymptotic wind velocity

\beq u_\infty \sim 2.1\Omega_0 r_0.
\label{uinf3}
\eeq

These can be used for comparison with observations.

\section{Disk Accretion Rate in PNe}

In order to produce a more detailed comparison of MHD disk winds
with PNe it is necessary to have a model for PNe accretion disks.
In particular it is necessary to know the accretion rate
$\dot{M}_d$, the inner disk radius $r_i$ as a function of time.

It is unlikely that an accretion disk could survive the long main
sequence lifetime of a PN central star. Thus, unlike YSOs and AGN,
accretion disks in PNe systems must form via binary interactions
Disks may form around secondaries via Roche lobe overflow or
accretion of the dense AGB wind (Morris 1987, Mastrodemos \&
Morris 1998). Such systems would be similar to symbiotic stars
(Corradi \ea 2000). Accretion disks could also form around the
primary after CE evolution and disruption of the secondary star
(Soker \& Livio 1994, Soker 1998, Reyes-Ruiz \& Lopez 1999).

Mastrodemos \& Morris (1998) carried out detailed simulations of
the first scenario.  Using a SPH method they modelled the
gravitational interaction of a dense AGB wind with a lower mass
companion.    They found steady accretion disks around a white
dwarf companion orbiting a AGB star with $\dot{M}_{agb} \approx
10^{-5}$ \mdot.  The ratio of $\dot{M}_d/\dot{M}_{agb} \approx .05
~-~ .005$ they found in their models is consistent with
expectations from basic theory (\cite{Frankea02})

Accretion disks may also form via disruption of the secondary
after CE evolution (Soker \& Livio 1994). This model implies a
finite lifetime for the disk as the mass reservoir of the
disrupted companion is slowly drained onto the primary.  A
description of disk formation in PNe has been given in
\cite{RRL99}.  Envelope ejection occurs via transfer of angular
momentum during which the secondary falls to a separation such
that it catastrophically overflows its Roche lobe and forms a disc
around the primary. There are a number of important constraints on
the properties of binaries which would lead to disk formation in
this way. \cite{RRL99} found that systems with a primary
consisting of an evolved AGB star with mass $M_* \approx 2.6 ~-~
3.6 ~M_\odot$, a low mass secondary ($\le 0.08$ \msol) and an
initial binary separation of $< 200 ~R_\odot$ may produce disks.
The AGB star will shed most of its mass during the common envelope
ejection, leaving a post-AGB stellar core surrounded by a thin
shell.

\cite{RRL99} find the disk accretion rate to evolve in time with
in a power-law manner.
\begin{eqnarray}
\dot{M}_d & = & \dot{M}_{do} \left( \frac{t}{1 ~yr} \right)^{-5/4}
{\rm{M}_\odot~yr^{-1}}
\label{rrlmdot}\\
\end{eqnarray}
\noindent Typical values of the scale is $\dot{M}_{do} = 10^{-3}$
\mdot.

\section{Disk Winds Models for PNe and pPNe}

We wish to understand if disk wind models can account for outflows
in PNe and pPNe and to characterize the parameters for the winds.
The momentum $\dot{\Pi} = \dot{M}_w u_w$ and energy $\dot{E} =
\frac{1}{2}\dot{M}_w u_w^2$ injection rates for PNe are easily
approximated. For "classic" PNe total mass of $M_{pn} \approx .1$
\msol ~must be accelerated to velocities of $u_{pn} \approx 40
~km/s$ in a timescale of order $\Delta t_{pn} \approx 10000 ~y$
(Note, here the subscript {\it pn} refers to observed properties
of the total nebula which reflects material swept-up by the wind
we model in this paper.) This gives $\dot{\Pi} = M_{pn}
u_{pn}/\Delta t_{pn} \approx 10^{27} ~g ~ cm/s^{2}$ and $\dot{E} =
M_{pn} u_{pn}^2/\Delta t_{pn} \approx 10^{34}~erg/s$.

Recall that \cite{Bujarrabalea01}) found high total outflow
momentum $10^{36} < \Pi/(g ~cm ~s^{-1}) < 10^{40}$ and total
outflow energy $10^{41} < E/(erg ~s^{-1}) < 10^{47}$ in an
extensive sample of pPNe. These values can be converted into
momentum and energy injection rates using an assumed injection or
"acceleration" timescale $\Delta t$, ($\dot{\Pi} = \Pi/\Delta t,
~L = E/\Delta t$.  Recall that the values for $\Pi$ and $E$ quoted
above can't be explained via radiation wind driving. Note also
that there is uncertainty about $\Delta t$, the injection
timescale, but most likely $\Delta t < 10^3 ~y$. The question
which then arises is: can energy and momentum budgets be met with
disk wind models?

The results of section 2 demonstrate that disk wind solutions are
sensitive to the location of footpoints for the flow $r_0$ and the
accretion rate $\dot{M}_d$. In what follows we use assume that the
inner edge of the disk extends to the stellar surface and use $r0
= r_i = r_*$.  We shall see that the accretion rate is the
parameter which becomes most important for obtaining solutions for
pPNe since the requisite high outflow momenta will require high
values of $\dot{M}_d$. One means of achieving high accretion rates
will be to use the model described by RRL99 where accretion rates
as high as a few times $10^{-4}$ \mdot are possible for short
periods as the disrupted companion's mass is fed onto the surface
of the primary.

{\bf PNe Solutions:} We first consider the case of a "classic"
PNe. In this case we would consider that the star which produces
the jet is a proto-WD with an AGB companion (Soker \& Rappaport
2000). Thus accretion rates of $\dot{M}_d \approx 10^{-6}$ are
resonable. To evaluate the expressions above we choose parameters
for a canonical PN with mass $M_s = 0.6$ \msol and a disk with
$\alpha = 0.1$ and $r_0/h_0 = 10$.

If we assume typical PNe central star parameters ($T_* = 10^5 ~K$,
$L_* = 5000 ~L_\odot$ such that $r_i = 1.64 \times 10^{10} ~cm$)
we find the following conditions for the wind from equations
\ref{mdot5} - \ref{uinf3} (note we use $u_w=u_\infty$),
\begin{eqnarray}
\dot{M_w} & = & 1\times10^{-7} ~\rm{M}_\odot~yr^{-1}
\left(\frac{\dot{M}_a} {10^{-6}
~\rm{M}_\odot~yr^{-1}}\right) \\
u_w & = & 1.25\times10^3 ~km/s
\left(\frac{M_*}{.6~\rm{M}_\odot}\right)^{1/2}
\left(\frac{r_0}{.23 ~\rm{R}_\odot}\right)^{-1/2}. \\
\end{eqnarray}
\noindent Thus using typical conditions for PNe central stars, the
scaling relations derived from the MHD equations yield disk wind
parameters well matched with observations.

{\bf pPNe Solutions:} While the mass loss rates and velocities are
known for PNe winds the situation for pPNe is not as clear.  In
general what is observed in pPNe is the total mass in the
outflows.  Mass loss rates must be inferred from the estimates of
pPNe acceleration timescales.  Velocities are also uncertain in
the sense that the winds themselves may not be observed directly
but only properties of swept-up material may be determined.

We assume a model post-AGB star with mass $M_s = 0.6$ \msol, a
temperature of $T_* = 10,000 ~K$ and a luminosity of $L_* = 5
\times 10^3$ \lsol ~ which, assuming a blackbody, yields a radius
of $r_* = 1.6 \times 10^{12} ~cm~= 23$ \rsol. Note that such a
star has an escape velocity of $u_{esc} = 98 ~km ~s^{-1}$

From our previous discussion it is clear that achieving the high
momentum input rates observed in pPNe via MCL disk wind models
will necessitate high accretion rates. Thus we adopt an
acceleration timescale of $\Delta t = 200 ~yr$ and an accretion
rate of $\dot{M}_d = 1 \times 10^{-4}$ \mdot. This value of
$\dot{M}_d$ is the $200$ year average of that found by RRL99 for
their case A ($\dot{M}_{do} = 1.6 \times 10^{-3}$ \mdot, equation
\ref{rrlmdot}).  Once again we choose a disk with $\alpha = 0.1$
and $r_0/h_0 = 10$.

Assuming $r_0 = r_*$ along with the other parameter values given
above, the key wind quantities $\dot{M}_w$, $u_w=u_\infty$ and
$L_m = L_w, \dot{E}$ can again be determined from equations
\ref{mdot5}, \ref{uinf3} and \ref{maglum},
\begin{eqnarray}
\dot{M_w} & = & 1. \times10^{-5} ~~\rm{M}_\odot~yr^{-1}~
\left(\frac{\dot{M}_d}{10^{-4} ~\rm{M}_\odot~yr^{-1}}\right) \\
u_w & \simeq & 146 ~km/s \left(\frac{M_*}{.6 ~\rm{M}_\odot
}\right)^{1/2}\left(\frac{R_i}{23 ~\rm{R}_\odot}\right)^{-1/2} \\
L_m & \simeq & 6.7 \times 10^{34} erg ~s^{-1}
\left(\frac{\dot{M}_d}{10^{-4} ~\rm{M}_\odot~yr^{-1}}\right)
\left(\frac{M_*}{.6 ~\rm{M}_\odot }\right) \left(\frac{R_i}{23
~\rm{R}_\odot}\right)^{-1}
\end{eqnarray}
Note that $L_w \Delta t \approx 10^{44}$, a value in the middle of
the range found by Burharrabal et al 2001.  Note also that the
solution above has $u_\infty \approx 1.5 u_{esc}$.  Since
$u_\infty \approx u_{esc}$ the higher velocity outflows seen in
some pPNe would require more disks around more compact central
sources.  We note here that the observed momenta and energy in the
outflows comes primarily from swept up circumstellar material.
Thus, as is the case in YSO molecular outflows (Bachiller 1996),
the energy and momentum budget of the disk driven outflow must be
sufficient to power the observed outflows via so-called prompt or
shock driven entrainment.

Given a model for the temporal history of the disk accretion, the
total energy and momentum for the outflows can be found. Replacing
$\dot{M}_d$ in equation \ref{rrlmdot} with $\dot{M}_d(t)$ from the
relations derived by \cite{RRL99} and integrating with respect to
$t$ gives
\begin{eqnarray}
E    & = & \int \frac{1}{2}\dot{M}_w u_\infty^2 dt \approx 1.3
\times10^{44} ~erg
~\left[1 - \left(\frac{1 ~yr}{t}\right)^{1/4})\right] \\
\Pi  & = & \int \dot{M}_w u_\infty dt \approx 1.8\times10^{37} ~g
~cm ~s^{-1}~ \left[1 - \left(\frac{1 ~yr}{t}\right)^{1/4}\right]
\end{eqnarray}
These results show that the MCL disk wind models can achieve both
energy and momentum injection rates {\it as well as} the total
energy and momentum required to account for many pPNe described by
Bujarrabal et al (2001). The total energy and momenta budgets we
find from these solutions fall well within the range of pPNe
outflows with momentum excesses. Taken together with our previous
calculations for "classic" PNe winds these results confirm the
predictions of \cite{BFW01} that magnetized disk winds can account
for much of the outflow phenomena associated collimated outflows
in the late stage of stellar evolution.

The results above indicate that collimated flows which form from
transient disks in the pPNe stage will appear as dense knots in
mature PNe flows. This may also serve to explain the presence of
so-called FLIERS (Fast Low Ionization Emission Regions: Balick \ea
1994) seen in some PNe. The mass loss rate in the winds derived
above rapidly decrease with time. Thus the bulk of the jet's mass
will lie near its head. As the material in the disk is accreted
onto the star the jet will eventually shut-off leaving the dense
knot to continue its propagation through the surrounding slow
wind.

When the star makes its transition to a hot central star of a PNe
its fast, tenuous spherical wind sweeps up a shell of the slow AGB
wind material.  The shell's expansion speed will typically be of
order $40 ~km/s$ and it will not catch up to the head of the jet.
Thus during the PNe phase the jet head will appear as a dense,
fast moving knot which should lie outside the PNe wind blown
bubble. We note that masses of FLIERs are estimated to be of order
$10^{-4} - 10^{-5} ~\rm{M}_\odot$ which is reasonable for the
models presented above. FLIER velocities can be lower than the
$\approx 100 ~km/s$ calculated above but deceleration of the jet
head will occur via interaction with the environment. We note also
that hydrodynamic simulations of PNe jets in which the jet ram
pressure decreases in time (as would occur for our model) show
characteristic patterns of backward pointing bow-shocks (apex
pointing back towards to the star, Steffan \& Lopez 1998).   If
such results are robust, the jets produced by disk winds in our
scenario above may also yield similar morphologies.

\section{Discussion and Conclusions}
While purely hydrodynamic models for PNe shaping provide an
adequate description for many large scale features seen in purely
morphological studies, jets and hourglass shaped bipolar nebulae
appear to strain their explanatory power.  In addition, the
momentum and energy associated with many pPNe appear to be orders
of magnitude larger than can be accounted for with radiation
driving from the central star (even when multiple scattering is
taken into account).  In this paper we have derived and applied
scaling relations derived from time-independent axisymmetric MHD
equations to the winds and wind-driven outflows in PNe and pPNe.

For "classic" PNe (a central star with $T_* >> 3\times10^4 ~K$) we
find magneto-centrifugal winds can account for typical observed
wind properties.  We find only a 1\% efficiency of accretion of an
AGB companion wind is required to produce reasonable results.

Our results for pPNe show that momentum excesses need not occur
for outflows driven by MCL winds.  While this is encouraging in
terms of finding a mechanism for driving pPNe outflows the
solutions require fairly high accretion rates ($ > 10^{-5}$
\mdot).  It is not clear if such conditions can be achieved with
the frequency required by observations. While solutions of
\cite{RRL99} yield accretion rates and time dependencies which
lead to the correct outflow momenta and energetics, their models
place fairly stringent limitations on the nature of the binaries
that form disks from disrupted companions. If accretion onto
undetected compact orbiting companions is invoked
(\cite{SokerRappaport01}) then higher values of $\dot{M}_d$ may
not be required.

It is worth noting that these models imply an outflow from both
the AGB wind and the jet.  The AGB wind can, in princple be
sculpted by its own magnetic forces or can be swept at later times
when a radiation-driven wind from the exposed core is initiated.
Thus this class models implies the possible mis-alignment between
the jet and the main body of the nebula. In Blackman, Frank \&
Welch (2001) a model for such multi-polar outflows was proposed in
which magnetically launched outflows are driven from the both the
disk and a non-aligned rapidly rotating AGB core. In this paper we
did not consider the AGB wind to be magnetized and considered the
field in the disk to be arise via a turbulent dynamo there. If the
AGB wind is magnetized then shaping could occur with an alignment
that is uncorrelated with that of the collimated disk wind. We
note that current MHD models of AGB winds however tend to support
the development of a torus (Matt et al 2001) rather than a jet.
Jet formation via initially weak fields in post-AGB winds have
been explored (Garcia-Segura et al 1999) however these models
require higher winds speeds than occur on the AGB. Finally it is
worth noting that recent simulation results by Matt, Blackman \&
Frank 2004 confirm that the exposed rapdily rotating magnetic core
model can produce well collimated outflows.  Regardless of these
cavets the point remains that collimated disk winds and AGB wind
systems can, in principle, explain multi-polar outflows.

We note that the presence of molecules in fast moving outflows
driven by winds, as proposed in our studies, is not new.  Studies
of YSO molecular outflows have revealed "fast molecular" material
moving at speeds of more than $25$ km/s, which is the nominal
dissociation speed for $H_2$ in J-shocks. A number of theoretical
proposals to resolve the issue currently exist including magnetic
precursors in J-shocks (Harigan et al 1989), high velocity
C-shocks (Smith \& Brand 1990), accelerating shocks (Lim 2003) and
the presence of clumped or inhomogeneous gas  (Hartquist \& Dyson
1987). Thus the behavior we see in proto-Planetary nebulae also
exists in YSOs where compelling evidence for MHD driven disk winds
already exists.  The question of how molecules can survive after
being swept-up in high velocity flows remains unanswered for pPNe
and YSOs but is currently an active area of research with a number
of competing models currently under investigation.

We note that a robust prediction of our models is the ratio of
wind mass loss rate to accretion rate, i.e.
$\dot{M}_{w}/\dot{M}_{a} \approx .1$. This is true for most MCL
disk wind models and can be seen as a target prediction which can
be explored observationally.

Thus the question which must now be addressed is can conditions
for either high accretion rate disks in pPNe or symbiotic type
accreting companions be made to embrace enough systems to account
for the statistics of \cite{Bujarrabalea01}.  There are many
uncertainties concerning the formation of disks via the disruption
of the companion after a CE phase.  The disruption of a secondary
may not really lead to an "accretion disk" around the primary due
to both tidal and hydrodynamical disruption.  The requirement that
material from the disrupted companion "push" trough the envelope
may not lead to a disk but rather a dense and clumpy expanding
torus.  A detailed discussion of the process however remains
outside the scope of our paper.  Readers wishing to consider the
viablity of these models are encouraged to review Soker \& Livio
(1994) and Ruiz-Reyes \& Lopez (1999).

There is an important difference between the two disk formation
scenarios discussed here.  In the case where AGB wind material is
captured to form a disk, one expects the abundances of the jet and
AGB outflows to be the same. In the disrupted companion scenario
the disk/jet will have formed from a different star in a different
evolutionary state.  Thus we would expect abundance differences
between the AGB and jet outflows.  It may, therefore, be
worthwhile for observers to search for abundance gradients between
different components of multipolar flows.

We note finally that this paper comprises a step beyond Blackman,
Frank \& Welch (2001) in establishing the efficacy of MHD
paradigms for pPNe/PNe in which strong magnetic fields play a role
in both launching and collimating the flows. Future models should
aempt to include more detailed description of the physics of MCL
disk launching in these systems including time dependent models.
(see for example (von Rekowski 2004)

\section*{ACKNOWLEDGMENTS}
Thanks to T. Lery, T. Gardiner, R. Pudritz, S. Matt, B. Balick, N.
Soker, R. Corradi, H. van Horn and J. Thomas for related
discussions. Support by NSF grants AST-9702484, AST-0098442, NASA
grant NAG5-8428, HST grant, DOE grant DE-FG02-00ER54600. and the
Laboratory for Laser Energetics.

{}

\end{document}